# The Keynesian Model in the General Theory: A Tutorial

Raúl Rojas
Freie Universität Berlin
January 2012

This small overview of the General Theory is the kind of summary I would have liked to have read, before embarking in a comprehensive study of the General Theory at the time I was a student. As shown here, the main ideas are quite simple and easy to visualize. Unfortunately, numerous introductions to Keynesian theory are not actually based on Keynes *opus magnum*, but in obscure neo-classical reinterpretations. This is completely pointless since Keynes' book is so readable.

**Introduction**

John Maynard Keynes (1883-1946) completed the *General Theory of Employment, Interest, and Money* [1] in December of 1935, right in the middle of the Great Depression. At that point, millions of workers in the US and Europe had been unemployed for years, and economic orthodoxy could not account for this "anomalous" situation. Keynes' *General Theory* tries to tackle exactly this problem. Keynes rejected classical theories based on the idea that production creates its own demand, that is, that the economy always recovers to full employment after a shock. Therefore, Keynes called his treatise the *General Theory* because he conceived classical doctrine as only a special case of a more complete approach:

> "The classical theorists resemble Euclidean geometers in a non-Euclidean world who, discovering that in experience straight lines apparently parallel often meet, rebuke the lines for not keeping straight (...) Yet, in truth, there is no remedy except to throw over the axiom of parallels and to work out a non-Euclidean geometry. Something similar is required to-day in Economics." (p. 16)

The axiom of the classical theorists, to which this passage refers, was that long-term unemployment, at its bottom, is voluntary, because workers always resist a reduction of wages. If production creates its own demand, and wages stay flexible, eventually all workers would be called, up to the point where their wages would equalize the marginal productivity of labor. The *General Theory* is thus the new required non-Euclidean approach for explaining long-term unemployment and the economic disequilibrium so rampant during the 1930s.

**Structure of the General Theory**

Keynes wrote the *General Theory* following a meticulous plan. The clean structure of the volume makes it easier to understand Keynes' model and the postulated relationships between dependent and independent variables. The manuscript is organized in six books:

- Book I: *Introduction*. This three chapters focus on the problem of unemployment, criticize the classical economists, and give an overview of the system to be developed.
- Book II: *Definitions and Ideas*. This is a methodological section justifying the choice of units and the definitions of economic terms such as income, savings, and investment.
- Book III: *The Propensity to Consume*. This is a core chapter where the principle of effective demand is thoroughly explained, and the investment multiplier is introduced.
- Book IV: *The Inducement to Invest*. Here investment is described as determined by expectations and the rate of interest. The quantity of money and liquidity preference, in turn, determine the interest rate. The General Theory is hence complete.
- Book V: *Money-Wages and Prices*. This is a kind of complementary section bringing the employment function, prices and wages into a functional relationship.
- Book VI: *Short Notes Suggested by the General Theory*. The chapter about the trade cycle in this section illustrates the full dynamics of the Keynesian model.

Paul Krugman has said that the GT is like a meal, with an appetizer at the beginning (Book I) and dessert at the end (Book VI) [2]. The main course is contained in the Books II, III, and IV. Book V is more of a complement, to make the theory air-tight against classical explanations of unemployment based on the level of money-wages and prices.

**The Principle of Effective Demand**

The main idea introduced in Book I is the *principle of effective demand*. Classical theorists assume that all production generates its own demand (Say's Law[1]). They equate production with income, and all income is used for acquiring all produced goods: supply and demand move in parallel until full employment is achieved. Keynes, however, makes a difference between production/supply, and demand: the aggregate supply function is the total price of goods produced by employing N persons. The aggregate demand is that demand for goods caused by employing also N persons. Both curves are not the same: the point where they meet is called the *effective demand*. "This is the substance of the General Theory".

In Chapter 4 (The Choice of Units), Keynes explains why he will refer to output in terms of aggregate prices. Since there are many different industries, referring to physical output makes aggregation tricky or impossible. Employment is an independent variable and skilled labor can be handled as a multiple of unskilled labor. "I propose, therefore, to make use of only two fundamental units of quantity, namely quantities of money-value and quantities of employment". In Input-Output-Analysis we would represent the economy today by a matrix with *n* sectors (*n* rows and *n* columns) which describes the flow of physical products from one sector to the other. Multiplying this matrix with the vector of unitary prices for the *n* sectors, we can compute the total supply of each sector, and the sum of the supply of all sectors would produce the aggregate supply for the economy. However, Input-Output-Analysis was still in its infancy and its methods are missing in the General Theory.

---

[1] „It is worthwhile to remark that a product is no sooner created than it, from that instant, affords a market for other products to the full extent of its own value", J.B. Say, *A treatise on political economy; or the production distribution and consumption of wealth*, 1803.

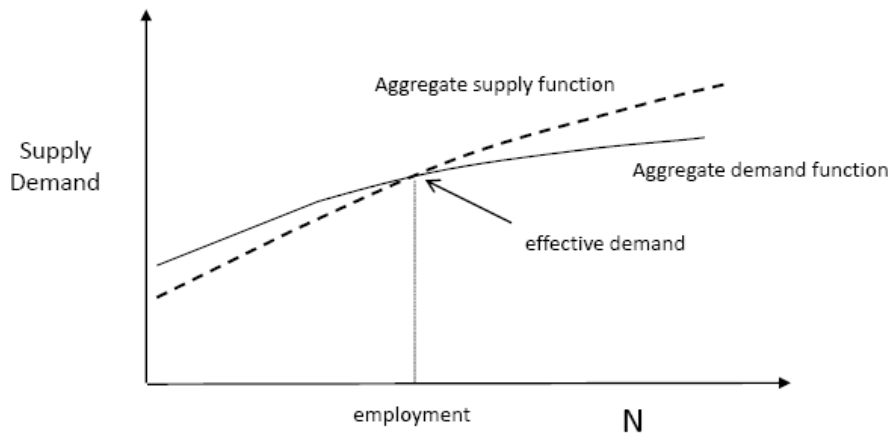

Fig. 1: The effective demand is the crossing point of aggregate supply and aggregate demand

The message to be developed in Books III and IV is that the effective demand can be too low so that full employment is not achieved. In Fig. 1, at low employment, there is more demand than supply (households have to spend their savings because the total wages are too low). Full employment requires a level of aggregate demand which is not available. Supply and demand meet therefore at an intermediate point, the effective demand. Aggregate demand can be too low due to insufficient consumption or insufficient investment, or both, the next problem to be explained.

**The propensity to consume**

Aggregate demand is composed of demand $C$ for consumption and demand $I$ for investment goods. Both are governed by different rules. In Keynes' opinion not all income $Y$ transforms automatically into demand $C+I$ for finished goods. The more income $Y$ a person obtains the less proportion he or she spends in direct consumption $C$ of goods. The consumption function bends down at higher levels of aggregate income (see Fig. 2).

In many appraisals of the Keynesian model, it is assumed that consumption is a linear function of income $Y$ of the form $C = C_0 + cY$, where $C_0$ is a constant and $c$ is the fraction of income devoted to consumption. In the General Theory there is no diagram of this function, and clear indications that Keynes did not consider consumption to be a linear function of income. He states clearly that "these reasons will lead, as a rule, to a greater *proportion* of income being saved as real income increases" (p. 97). That is, the constant $c$ is lower, the higher $Y$ is.

The factors determining the level of consumption are objective and subjective. Objective factors are, for example, wage changes, changes in net income, time discounting (that is, if it is better to buy something now or later), as well as future income expectations. The subjective factors are listed at the beginning of Chapter 9 and reduce to precaution, foresight, calculation, improvement, independence, enterprise, pride and avarice, as well as their opposites such as enjoyment, shortsightedness, generosity, miscalculation, ostentation, and extravagance (p. 108).

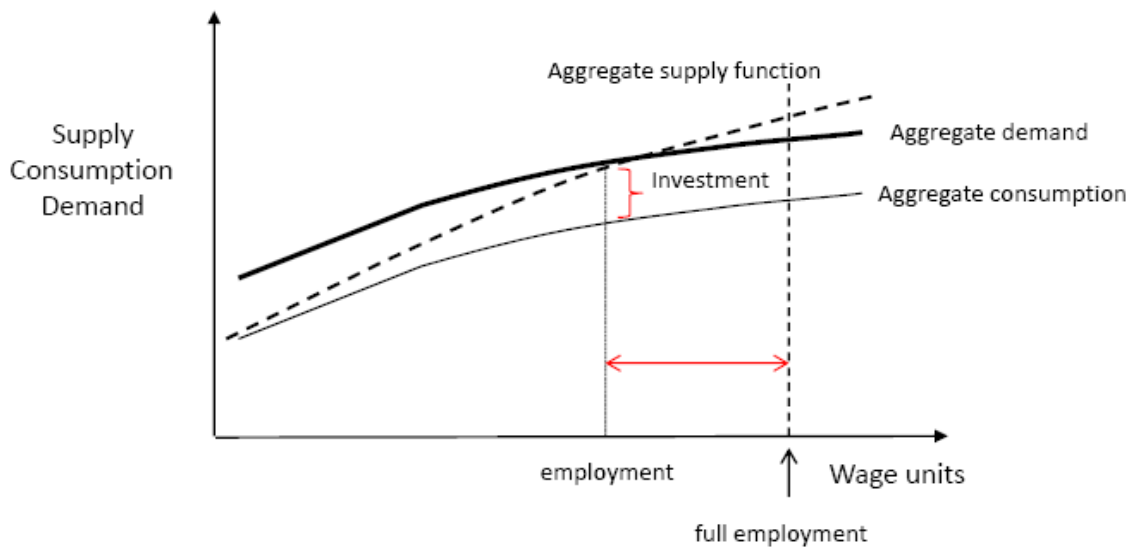

Fig. 2: Aggregate demand and aggregate supply do not necessarily cross at full employment level

Fig. 2 shows a hypothetical consumption function and a fixed level of investment I. Notice that the horizontal axis is now expressed in terms of wage units, so that the more employment we have, the more wages are paid. The horizontal axis provides us with a variable which can measure how much income is in the hands of society. The effect of a fixed investment I is to displace the consumption function up, producing an aggregate demand which intersects the aggregate supply function. In this example, equilibrium has been found at less than full employment. The later could only be achieved increasing the investment I, or raising the consumption function. Keynes considers the consumption function rather stable, so that filling the employment gap can only be done by increasing the investment.

Keynes considers thus the special concave form of the consumption function as a given: "The fundamental psychological law, upon which we are entitled to depend with great confidence both a priori from our knowledge of human nature and from the detailed facts of experience, is that men are disposed, as a rule and on the average, to increase their consumption as their income increases, but not by as much as the increase in their income."

**Investment and the multiplier**

In conditions where full employment has not been reached, a higher level of investment can fill the gap between the consumption and the supply function. In Fig. 3 now the horizontal axis represents spendable income and the vertical axis total demand. If income equals demand, the economy is in equilibrium (that is, all points in the 45 degree diagonal represent possible levels of equilibrium for different total incomes). Fig. 3 shows that increasing investment from $I_1$ to $I_2$ displaces aggregate demand from $C+I_1$ to $C+I_2$ so that a new equilibrium, at a higher level of income and employment, can be achieved. The interesting point here is that the increase ΔY in total income is higher than the increase ΔI of the investment. The red line shows the path to a new equilibrium: the higher investment leads to more demand and thus higher income. Higher income translates into higher consumption at the increased investment level. This leads again to higher income an so on, up to the new point where aggregate supply and demand are in equilibrium again. The income increase is a multiple of the increase in investment. Enter the multiplier.

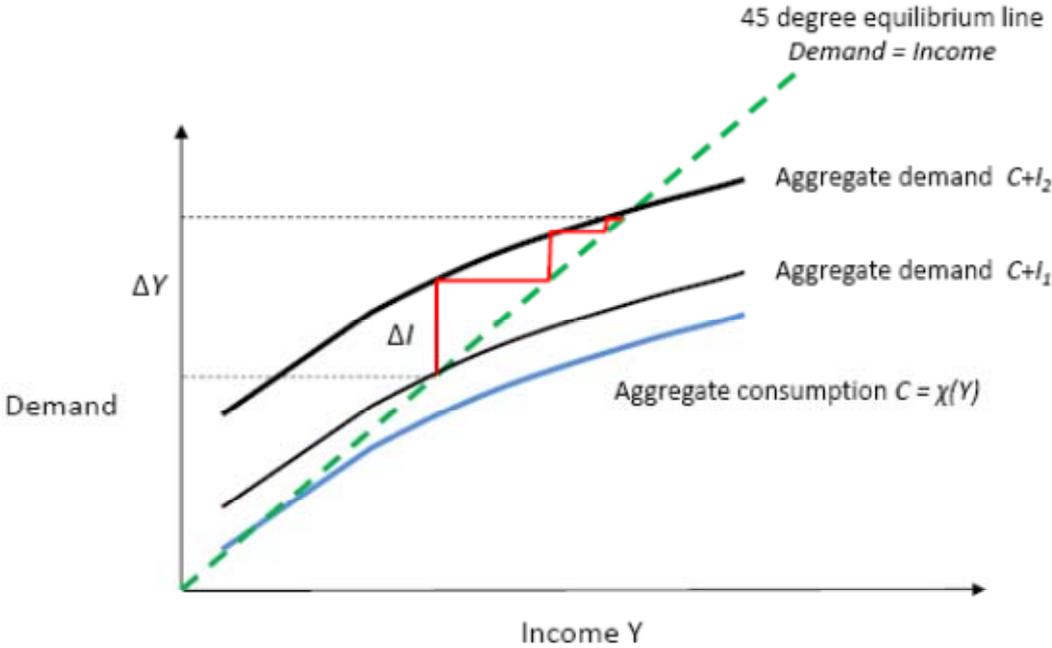

Fig. 3: An increase ΔI=$I_2$-$I_1$ in investment produces an even higher increase ΔY in income

The derivation of an expression for the multiplier is easy. The variables are income $Y$, consumption $C$ and investment $I$. We know that $Y = C + I$ and assuming that consumption $C$ is a fraction $c$ of Y we get $Y = cY + I$. Considering increments we obtain:

$$\Delta Y = c\Delta Y + \Delta I,$$

and therefore

$$\Delta Y = \frac{\Delta I}{1 - c} = k\Delta I$$

The number $k$ is called the "investment multiplier" or alternatively the "income multiplier". Since $c$ is lower than 1, the multiplier is higher than 1. Given Keynes "fundamental psychological law" the multiplier has to be always larger than 1.

Notice that this derivation is valid for small increments of the investment. For larger increments the assumption that $c$ is constant does not hold and a more complex expression would have to be derived.

The investment multiplier is thus a wonderful thing, most of all when society has reached a level of income with a low propensity to consume. If the private sector is not investing enough, the state can take the initiative and increase the level of investment. The effects spill over to all sectors of society: "If the Treasury were to fill old bottles with banknotes, bury them at suitable depths in disused coalmines which are then filled up to the surface with town rubbish, and leave it to private enterprise on well-tried principles of laissez-faire to dig the notes up again (...), there need be no more unemployment and, with the help of the repercussions, the real income of the community, and its capital wealth also, would probably become a good deal greater than it actually is. It would, indeed, be more sensible to build houses and the like; but if there are political and practical difficulties in the way of this, the above would be better than nothing" (p. 129).

**Investment and the interest rate**

Keynes rejects the classical theory according to which, the interest rate is determined by the equilibrium between supply and demand of loanable funds. In Keynes' model the interest rate is determined by the money supply and the curve of liquidity preference (see next section). Whatever the interest rate, for each amount of investment first the most productive capital goods will be bought, and then the less productive ones. The "schedule of the marginal efficiency of capital" is a decreasing function which relates total investment $I$ to interest rate $r$. When the interest rate is high, only a few investment opportunities are profitable. When the interest rate is low, more money can be invested since more options are available.

The schedule of marginal efficiency is in part psychological, since it is determined by the *expected* future rewards of an investment. That is, economic pessimism can decrease investment for a given interest rate. Economic optimism can raise investment without any change of the interest rate. This is illustrated in Fig. 4 where the different expected schedules for the marginal efficiency of capital displace upwards with increasing optimism. Given these expectations and the interest rate, the level of investment gets determined.

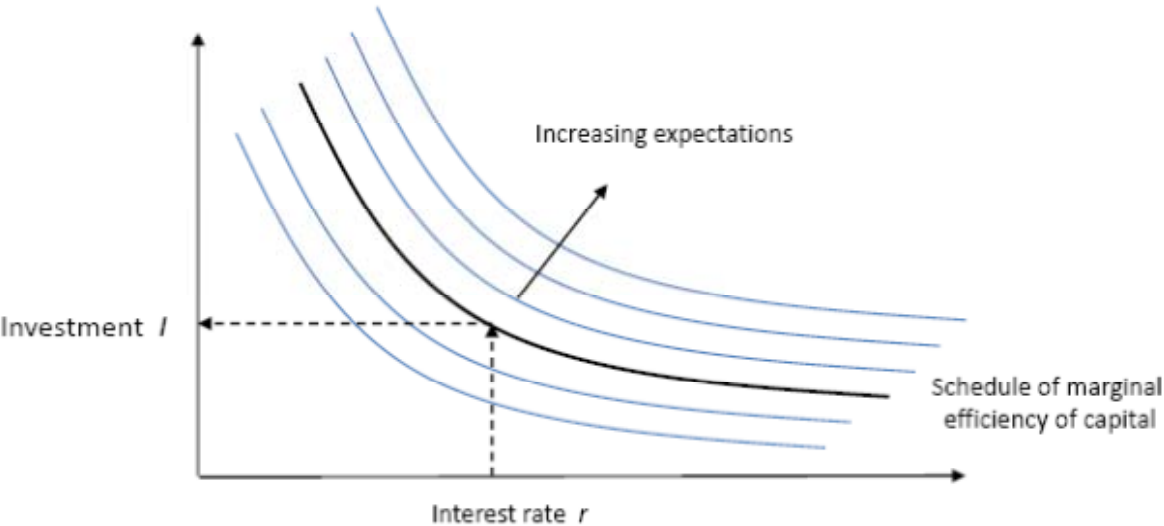

Fig. 4: The expected schedule of the marginal efficiency of capital and the interest rate $r$ determine the investment $I$

However, investors can err on the side of caution if their expectations are too low. Here the state can intervene and give a little nudge to the economy: "For my own part I am now somewhat skeptical of the success of a merely monetary policy directed toward influencing the rate of interest. I expect to see the state, which is in a position to calculate the marginal efficiency of capital goods on long views and on the basis of the general social advantage, taking an ever greater responsibility for directly organizing investment; since it seems likely that the fluctuations in the market estimation of the marginal efficiency of different types of capital, calculated on the principles I have described above, will be too great to be offset by any practicable changes in the rate of interest."

**The money supply, liquidity preference, and the interest rate**

For Keynes, the interest rate is what investors are ready to pay someone for his "liquidity". There are many reasons why a person would prefer to keep money unused: for future use, in consumption or investment, or as a precaution against unforeseen events. There can also be speculative motives. Keynes calls the cash held for future transactions or precaution $M_1$, and the cash held for speculative motives $M_2$, then

$$M = M_1 + M_2 = L_1(Y) + L_2(r)$$

where $L_1$ is a function of Y and the speculative liquidity preference is a function of the interest rate r.

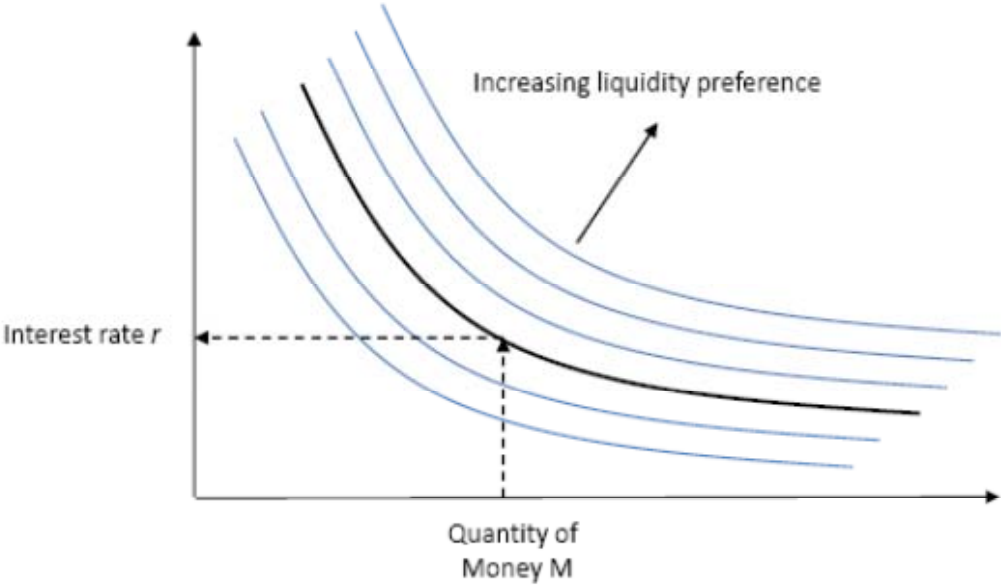

Fig. 4: The current liquidity preference curve and the money supply determine the interest rate r

Fig. 4 shows the way that the curve of liquidity preference determines the rate of interest. The various curves denote different levels of income and different psychological levels of liquidity preference (for example during a crisis). But in general, the more money that is available, the lower the rate of interest, and vice versa. Therefore, given a certain liquidity preference and a certain quantity of money, the rate of interest is fixed as shown in the figure. Keynes points out: "Liquidity-preference is a potentiality or functional tendency, which fixes the quantity of money which the public will hold when the rate of interest is given; so that if r is the rate of interest, M the quantity of money and L the function of liquidity-preference, we have $M = L(r)$. This is where, and how, the quantity of money enters into the economic scheme."

**Keynes economic machine**

The four figures above capture the essentials of Keynes' General Theory. At the end of Book IV, in Chapter 18, Keynes proceeds to summarize the variables:

- The independent variables are: propensity to consume, the schedule of the marginal efficiency of capital, and the rate of interest (they are capable of further analysis)
- The dependent variables are the volume of employment and the national income.

And Krugman's own summary [2] is in the form of bullet points:

"• Economies can and often do suffer from an overall lack of demand, which leads to involuntary unemployment
• The economy's automatic tendency to correct shortfalls in demand, if it exists at all, operates slowly and painfully
• Government policies to increase demand, by contrast, can reduce unemployment quickly
• Sometimes increasing the money supply won't be enough to persuade the private sector to spend more, and government spending must step into the breach."